\documentstyle[12pt,epsf,cite]{article}
\begin{document}
%\large
%\begin{titlepage}
\begin{center}
{\Large\bf New Evidence for
 Supernarrow Dibaryons Production in $pd$ Interactions}\\
\vskip 5mm

%\author{
L.V. Fil'kov$^{1)}$\footnote{
E-mail: filkov@sci.lebedev.ru},
V.L. Kashevarov$^{1)}$, 
E.S.Konobeevski$^{2)}$\footnote{E-mail: konobeev@sci.lebedev.ru}, 
M.V. Mordovskoy$^{2)}$,\\ 
S.I. Potashev$^{2)}$, 
V.A. Simonov$^{2)}$, 
V.M. Skorkin$^{2)}$, 
S.V. Zuev$^{2)}$

\vskip 5mm
 $^1$ Lebedev Physical Institute, RAS, Leninsky Prospect 53, 117924 Moscow, Russia \\
$^2$  Institute for Nuclear Research, RAS, 60-th October Anniversary Prospect 7a,
117312 Moscow, Russia\\ 

\end{center}

\vskip 5mm
\begin{abstract}
The analysis of new experimental data, obtained at the Proton Linear
Accelerator of INR, with the aim to search for
supernarrow dibaryons in the $pd\to ppX_1$ and
$pd\to pdX_2$ reactions is presented. Narrow peaks with an 
experimental width of 5 MeV at masses 
of 1904$\pm 2$, 1926$\pm 2$, and 1942$\pm 2$ MeV have been observed
in missing mass $M_{pX_1}$ spectra.
In the missing mass $M_{X_1}$ spectra, the peaks
at $M_{X_1}=966\pm 2$, 986$\pm 2$, and 1003$\pm 2$ MeV
have been found. The analysis of the data obtained leads to the
conclusion that the observed peaks in $M_{pX_1}$ spectra are most likely 
supernarrow dibaryons, 
the decay of which into two nucleons is forbidden by the Pauli exclusion
principle. An alternative interpretation of the spectra by assuming a
decay of the supernarrow dibaryons in "exotic baryon states" with masses
$M_{X_1}$ is discussed.\\
\end{abstract}

\vskip 1mm
{\bf PAKS}: 13.75.Cs, 14.20.Pt, 12.39.Mk
%\maketitle
%\end{titlepage}
\newpage
\section{Introduction}

The experimental search for dibaryons continues since over 20 years
(for recent reviews see \cite{troy,tat1}).
The dibaryons considered couple usually to two nucleons and reflect their
symmetry. Therefore, such
dibaryons have estimated 
decay widths from several MeV up to a few hundreds MeV. 
However, their production cross sections in reactions involving nucleons
and pions are small and the spectra are dominated by a large background.
This has lead to contradictory
results and the dibaryons decaying directly into two nucleons are not 
unequivocally established.

We consider a new class of dibaryons -- supernarrow dibaryons
(SNDs), the decay of which into two nucleons
is forbidden by the Pauli exclusion principle
\cite{mul,fil1,fil2,ger}. Such dibaryons with masses
$M<2m_N+m_{\pi}$, where $m_N$($m_{\pi}$) is the nucleon (pion) mass, 
can decay into two nucleons, by emitting a third particle, normally a photon. 
The decay widths
of these dibaryons are predicted to be $\le 1$ keV \cite{fil2}.

In the framework of the MIT bag model, Mulders {\it et al.} \cite{mul} calculated
the masses of different dibaryons, in particular, $NN$-decoupled dibaryons.
These dibaryons $D(T=0;J^P =0^{-},1^{-},2^{-};M=2.11$ GeV) and
$D(1;1^{-};2.2$ GeV) correspond to the states $^{13}P_J$
and $^{31}P_1$ forbidden in the $NN$ channel.
However, the masses of
these dibaryon exceed the pion production threshold.
Therefore, these dibaryons can decay into the $\pi NN$ channel and
their decay widths are larger than 1 MeV. 

Using the chiral
soliton model, Kopeliovich \cite{kop} predicted that the masses of
$D(T=1,J^P=1^+)$ and $D(0,2^+)$ $NN$-decoupled
dibaryons exceeded the two nucleon mass by 60 and 90 MeV, respectively.
These values are lower than the pion production threshold.   

In the framework of the canonically quantized biskyrmion model
Krupovnickas {\it al.} \cite{riska} obtained an indication 
of the existence of one dibaryon with J=T=0 and two dibaryons with J=T=1 with
masses smaller than $2m_N+m_{\pi}$.

All these results are model calculations of very limited predictive power
and provide just a motivation to look for SNDs. In the following we
summerize the experimental attempts to look for such dibaryons so far.

In ref. \cite{khr} dibaryons with exotic quantum numbers were searched for
in the process $pp\to pp\gamma\gamma$. The experiment was performed with
a proton beam from the JINR phasotron at an energy of about 216 MeV. The
energy spectrum of the photons emitted at $90^{\circ}$ was measured and
showed two peaks. This behavior 
of the photon energy spectrum was interpreted as a signature of an exotic
dibaryon resonance with a mass of about 1956 MeV and possible isospin
$T=2$. However, the results of ref. \cite{str} make
the possibility of a production of dibaryons with $T=2$ in this reaction
questionable. So, additional careful studies of the
reaction $pp\to pp\gamma\gamma$ are needed to more correctly understand
the nature of the observed state.

An analysis of the Uppsala proton-proton bremsstrahlung data \cite{cal} 
looking for the presence of a dibaryon in the mass range
from 1900 to 1960 MeV gave only upper limits of 10 and 3 nb for   
the dibaryon production cross section at proton beam energies of 200 and
310 MeV, respectively.
This result agrees with the expected values of the cross
section obtained at the conditions of this experiment in the framework  
of the model of dibaryon production and decay suggested in ref. \cite{prc} 
and does not contradict the data of ref. \cite{khr}.

In our previous studies \cite{izv,ksf,yad,prc} we investigated the
reaction $pd\to pX$ in order to search for SNDs. 
In this process SNDs can be produced only if
the nucleons in the deuteron overlap sufficiently, such that a six-quark 
state with deuteron quantum numbers can be formed. Then, an interaction of
a meson or another particle with this state can change its quantum
numbers so that a metastable state is formed.
The experiment was
carried out at the proton beam of the Linear Accelerator of INR using
the two-arm spectrometer TAMS. As was shown in ref. \cite{yad,prc},
the nucleons and the deuteron from the decay of SND into $\gamma NN$
and $\gamma d$ have to be emitted into a narrow angular cone with respect to
the direction of the dibaryon. If a dibaryon
decays mainly into two nucleons, then the expected  angular cone of the
emitted nucleons is about $50^{\circ}$. Therefore, a detection
of the scattered proton in coincidence with the proton (or the deuteron)
from the decay of particle $X$ at correlated angles allowed to suppress
effectively the contribution of the background processes and to increase
the relative contribution of a possible SND production. As a result,
two narrow peaks in missing mass spectra have been observed at
$M=$1905 and 1924 MeV with widths equal to the experimental resolution
($\sim 3 MeV$) and with 4.8 and 4.9 standard deviations (SD),
respectively. The analysis of the angular distributions of the charged
particles ($p$ or $d$)
from the decay of particle $X$ showed that the peak found at 1905 MeV
most likely corresponds to a SND with isotopic spin equal to 1.
In ref. \cite{prc} arguments were presented that the resonance at
$M=$1924 MeV could be SND, too.

\section{Experiment}

In this paper we present a new study of the $pd\to pX$ reaction
at the Linear Accelerator of INR with 305 MeV proton beam using the
spectrometer TAMS
where the proton and deuteron from the decays $X\to pX_1$ and 
$X\to dX_2$ are measured in coincidence
with the scattered proton.

In the experiment, CD$_2$ and $^{12}$C were used as targets.
The scattered proton was detected in the left arm
of the spectrometer TAMS at the angle $\theta_L=70^{\circ}$. The second
charged particle (either $p$ or $d$) was detected in
the right arm by three telescopes located at $\theta_R=34^{\circ}$,
$36^{\circ}$, and $38^{\circ}$.
%These angles were shifted by $1^{\circ}$
%in comparison with the conditions of our previous experiment,
%that allowed to get an additional information about an angular distribution
%of the charged particles from the decay of the dibaryons
%in question.

A trigger was generated by four-fold coincidences of the two $\Delta E$
detector signals of the left arm combined with those of any telescope of
the right arm. The events contained information about time-of-flights
and full energies of two particles detected in coincidence in
the left and right arms of the spectrometer. The energy resolution was
4 MeV (7 MeV) for the left (right) arm. A time-of-flight resolution better 
than 0.5 ns was achieved. Each valid event including two time-of-flights and
two energies were stored event by event and then analyzed off line.

An off-line identification of different particles was performed by means
of their energies $E$ and time-of-flights $t$. In this way, protons and other
charged particles were identified via characteristic loci observed in
two dimensional diagrams of time-of-flight versus energy. As an example
such an experimental $E-t$ distribution for the right arm detector
at $\theta_R=38^{\circ}$ is
displayed in Fig. 1. This figure shows the band of proton
events, presenting
the dependence of the proton time-of-flight on its energy, and a spot
corresponding to the deuteron from the elastic $pd$ scattering.

Several software
cuts have been applied to the mass spectra. The energy of the
scattered proton was limited by an interval of $50<E_2<150$ MeV,
that corresponded to the interval of measured dibaryon masses
$1980 >M_{pX_1}\; (M_{dX_2}) > 1860$ MeV/c$^2$.
In order to suppress the background from the reaction
$p\; ^{12}C\to dY$, where $Y$ stands for the unobserved particles, 
in $M_{dX_2}$ spectra, we omitted events with a 
deuteron energy higher than the energy of the elastically scattered deuteron 
in the $pd\to pd$ reaction at the
given angles. For the $M_{pX_1}$ spectra the energy interval for the
proton from the SND decay was determined by the kinematics of the
decay of $X$ into the $\gamma NN$ channel. This interval was equal to
$50<E_p<100$ MeV. Such a cut allows to suppress essentially the contribution
from the reaction $pd\to ppn$. The main contribution to this reaction 
at the energy under consideration is
given by one-nucleon pole diagrams when one of the nucleons is a
spectator. These diagrams give the maximum contribution when the kinetic energy
of the spectator nucleon approaches zero. 

Missing mass spectra of $M_{pX_1}$ and $M_{dX_2}$ were determined with
the expression
\begin{equation}
M^2_{pX_1(dX_2)}=m^2_d+2m^2_p+2m_d(E_i^t-E_f^t)-2E_i^tE_f^t+2p_ip_f\cos\theta_L
\end{equation}
with the additional condition that the proton from the $X\to pX_1$ and the
deuteron from the $X\to dX_2$ decay were
detected in the right arm detector of the TAMS spectrometer. 
In this expression $m_{p(d)}$ is the proton (deuteron) mass, $E_i^t$
and $E_f^t$ are the total energies and $p_i$ and $p_f$ are the momenta of the
incident and scattered protons, respectively.

The missing mass $M_{dX_2}$ spectrum of the reaction $pd\to p dX_2$
for the angle $\theta_R=38^{\circ}$
(Fig. 2) has a peak at the deuteron mass. This peak corresponds to the
elastic $pd$ scattering. A measurement of this reaction at different
angles of the right and left arms was used to calibrate
the spectrometer.
The over all mass resolution of the spectrometer was $\sim 5$ MeV, 
the angular resolution $\sim 1^{\circ}$.

\section{Results and Discussions}

We first consider the $pd\to pX\to ppX_1$ reaction. 
Figs. 3a-3c depict the experimental missing mass $M_{pX_1}$ spectra
obtained with the CD$_2$ and $^{12}$C targets,
where (3a), (3b), and (3c) correspond to a detection of the second
proton in the right arm detector at
$\theta_R=34^{\circ}$, $36^{\circ}$, and $38^{\circ}$, respectively.
The background in these spectra are interpolated by polynomials.
Three peaks at $M_{pX_1}=1904\pm 2$, 1926$\pm 2$, and 1942$\pm 2$ MeV are
clearly visible in these spectra.
The first two of them confirm the values of the dibaryon mass obtained
by us earlier \cite{izv,ksf,yad,prc}, whereas the resonance at 1942 MeV 
is a new one. This additional resonance is due to the different kinematical
coverage of the new experiment. In this experiment we could decrease
the detection threshold of the scattered proton down to 50 MeV giving
us a broader acceptance in the missing mass. 

The experimental missing mass spectra, obtained with the carbon target,
are rather smooth. This
smoothness is caused by both, an essential increase of the contribution of
background reactions in the interaction of the proton with the carbon
and the Fermi motion of the nucleons in the nucleus. The latter increases
essentially the angular cone of the emitted nucleons. In consequence,
it is not possible to see peaks of SNDs  
on the carbon target.
As the experiment with
the carbon target resulted in the rather smooth spectra, all structures,
appearing in the experiment with the CD$_2$ target, have to be attributed
to an interaction of the proton with the deuteron.

The experimental spectra in Figs. 3a-3c are compared with the prediction
of the theoretical model of SNDs $D(T=1,J^P=1^{\pm})$ production 
constructed in the one meson exchange approach \cite{prc}
and normalized to the values of peaks in Fig. 3a.
Our calculation for the isovector SND $D(T=1,J^P=1^{\pm})$ with the
mass $M$=1904 MeV
showed that the contributions of such a dibaryon to the spectra at 
angles $34^{\circ}$, $36^{\circ}$, and $38^{\circ}$ must
relate as $1:0.92:0.42$. For the isoscalar SND $D(0,0^{\pm})$ one obtains
$1:0.95:0.67$.
The biggest contribution of the SND with $M=1926$ MeV
is expected at $\theta_R=30^{\circ}\div 34^{\circ}$. However, the angles
$30^{\circ}\div 33^{\circ}$ were not
investigated in this work. The calculation of the ratio of the contributions
to the missing mass spectra at the angles
$34^{\circ}$, $36^{\circ}$, and $38^{\circ}$ gave $1:0.85:0.34$ for $T=1$
and $1:0.71:0.46$ for $T=0$.

Nucleons from the decay of the SND with $M=1942$ MeV 
have a wider angular distribution due to the the higher mass and
consequently the higher transverse momenta in the decay $X\to pX_1$
according to the calculations 
with a maximum in the region of $26^{\circ}\div 32^{\circ}$.
In the region of the angles under consideration here,
the contributions of the $D(1,1^{\pm})$ are expected to behave as
$1:0.6:0.2$. For the SND $D(0,0^{\pm})$ we have $1:0.78:0.55$.

All these predictions for the SNDs are in agreement with our
experimental data within the errors. However, the analysis of
the reaction $pd\to ppX_1$ only in the considered angular range
does not allow to determine an isotopic spin of the SNDs.

If the observed states are $NN$-coupled dibaryons decaying
mainly into two nucleons then the expected angular
cone size of emitted nucleons must be more than $50^{\circ}$.
Therefore, their contributions to the missing mass spectra in
Fig. 3a-3c would be nearly constant
and would not exceed a few events, even assuming that
the dibaryon production cross section is equal to that of elastic $pd$
scattering ($\sim 40 \mu$b/sr).
Hence, the peaks found most likely correspond to SNDs.

The missing mass $M_{dX_2}$ spectrum of the reaction 
$pd\to pX\to pdX_2$, for the sum of angles $\theta_R=34^{\circ}$ and 
$36^{\circ}$ is shown in Fig. 4.
As seen from this figure, the reaction $pd\to pdX_2$ gives a very
small contribution to the production of the dibaryons under study.

On the other hand, it is expected \cite{yad,prc} that isoscalar SNDs
contribute mainly
in the $\gamma d$ and isovector SNDs in the $\gamma NN$ channels.
As the main contribution of the indicated dibaryons is observed in $pX_1$
channel, it is possible to suppose that $X_1\to\gamma n$ and
all found states are isovector SNDs.
A more precise conclusion about the value of the  isotopic spin of the
observed SNDs could be obtained by the study of the reaction
$pd\to npX_1$. The experiment on the SND photoproduction in the
process $\vec{\gamma}d\to \pi^{\pm}D$ with polarized photons could separate
the $D(T=1,J^P=1^+,S=1)$ and $D(1,1^-,0)$ states \cite{alek} where
$S$ is the spin of the SND.

The sum of missing mass spectra of the reaction $pd\to ppX_1$ over
angles $\theta_R=34^{\circ}$ and $36^{\circ}$, where the contribution of
the SNDs is maximum,
is presented in Fig. 5a. This spectrum was fitted by a
polynomial for the background plus Gaussians for the
peaks. The number of standard deviations (SD) is then determined from 
this spectrum as
\begin{equation}
\frac{N_{eff}}{\sqrt{N_{eff}+N_{back}}}
\end{equation}
where $N_{eff}$ is the number of events above the background curve and
$N_{back}$ is the number of events below this curve.
Taking nine points for each peak, we have 6.0, 7.0, and 6.3 SD
for the resonances at 1904, 1926, and 1942 MeV, respectively. The widths of
these resonances are equal to the experimental resolution of $\sim 5$ MeV.

An additional information about the nature of the observed states
has been obtained by studying the missing mass $M_{X_1}$ spectra of the
reaction $pd\to ppX_1$.
If the state found is a dibaryon decaying mainly into two nucleons then
$X_1$ is a neutron and the mass $M_{X_1}$ is equal to the neutron mass
$m_n$. If the value of $M_{X_1}$, obtained from the experiment, differs
essentially from $m_n$ then $X_1\to\gamma n$ and we have the additional
indication that the observed dibaryon is SND.

%A simulation of the missing mass spectra for the reaction $pd\to ppX_1$,
%where $pX_1$ are the decay products of the SNDs
%with the masses 1904, 1926, and 1942 MeV, gave peaks
%at $M_{X_1}$=965, 987, and 1003 MeV, respectively.

The simulation of the missing $M_{X_1}$ mass spectra for the reaction
$pd\to ppX_1$ has been performed assuming that the SND decayed as
$SND\to\gamma d(^{31}S_0)\to \gamma pn$ through the two-nucleon singlet
state $^{31}S_0$ \cite{fil2,prc}. Such a decay is characterized in the
rest frame by a narrow peak near the maximum photon energy in the
probability distribution of the dibaryon decay over an emitted photon 
energy. As a result of this simulation, in the missing
$M_{X_1}$ spectra three narrow peaks at $M_{X_1}= 965$, 987, and 1003 MeV
have been predicted. These peaks correspond to
the isovector SNDs with the masses 1904, 1926, and 1942 MeV, respectively.
The result of this simulation without the influence of 
the detector resolution is shown in 
Fig. 6. Fig. 6a shows the $M_{X_1}$ mass spectrum when the proton from
the SND decay $SND\to \gamma pn$ is emitted into the whole allowed angular
region.
%For the isoscalar SNDs such peaks would be a little wider.
The acceptance of the detectors cuts into the widths of these peaks.
The missing $M_{X_1}$ mass spectra, taking into account the 
acceptance of the detectors at the angles 34$^{\circ}$ and 36$^{\circ}$, 
is presented in Fig. 6b. The experimental resolutions widen these peaks
to $6-8$ MeV.

Fig. 5b depicts the sum of the missing mass $M_{X_1}$ spectra  
%obtained from the experiment
for the angles $\theta_R=34^{\circ}$ and $36^{\circ}$.
As is seen from this figure, besides the peak at the neutron mass,
which is caused by the process $pd\to ppn$, peaks are
observed at $966\pm 2$, $986\pm 2$, and $1003\pm 2$ MeV. 
%These values of $M_{X_1}$ coincide with
%the ones obtained  from the simulation and differ essentially from
%the value of the neutron mass (939.6 MeV). 
Hence, for all states under
study, we have $X_1\to\gamma n$ in support of the assumption that the
dibaryons found are SNDs.

It should be noted that the peak at
$M_{X_1}=1003\pm 2$ MeV corresponds to the resonance found in ref. 
\cite{tat2} and was attributed to an exotic baryon state $N^*$ below the
$\pi N$ threshold.
In that work, the authors found altogether
three such states with masses 1004, 1044, and 1094 MeV.
This is not in contradiction to our interpretation, since in principle
SND could decay according to $X\to pN^*$.
The possibility of the
production of $NN^*$-coupled dibaryons was considered in \cite{tat1}.

If the exotic baryon
states decay into $N^*\to\gamma N$ then they would contribute
to the Compton scattering on the nucleon. However, the analysis
\cite{lvov} of the existing experimental data on this process excludes
$N^*$ as intermediate states in the Compton scattering on the nucleon.

In ref. \cite{kob} it was assumed that these states belong to totally
antisymmetric $\underline{20}$-plet of the spin-flavor $SU(6)_{FS}$ group.
Such a $N^*$ can transit into
a nucleon only if two quarks from the $N^*$ participate in the
interaction \cite{feyn}. Then the simplest decay of $N^*$ with the masses
1004 and 1044 MeV would be $N^*\to \gamma\gamma N$.
This conjecture could be checked, in particular,
by studying the reactions $\gamma p\to\gamma X$ or $\gamma p\to\pi X$
at a photon energy close to 800 MeV.

Taking into account the mentioned connection between the SNDs
and the resonancelike states $X_1$,
it is possible to assume that the peaks, observed in
\cite{tat2} at 1004 and 1044 MeV, are not exotic baryons, but
they are the resonancelike states $X_1\to\gamma n$
caused by the existence and decay of the SNDs with the masses 1942
and 1982 MeV, respectively.
Such $X_1$ would be not real resonances and cannot give contribution to the
Compton scattering on the nucleon.

\section{Conclusions}

The result of the study of the reaction $pd\to ppX_1$ is that 
three narrow peaks at 1904, 1926, and 1942 MeV have been observed in
the missing mass $M_{pX_1}$ spectra.  The analysis of the angular distributions of
the protons from the decay of the $pX_1$ states and the data of the reaction $pd\to
pX\to pdX_2$ showed that the peaks found can be explained as a manifestation of the
isovector SNDs, the decay of which into two nucleons is forbidden by the Pauli
exclusion principle.  The observation of the peaks in the missing mass $M_{X_1}$
spectra at 966, 985, and 1003 MeV is an additional confirmation that the dibaryons
found are the SNDs. 

\section*{Acknowledgment}

The authors thank L.V. Kravchuk and V.M. Lobashev for support of this
work and the team of the Linear Accelerator for the help in performance
of the experiment. The authors thank Th. Walcher for carefully reading
the manuscript and useful remarks.
The authors thank also J. Friedrich and B.M. Ovchinikov
for the active discussion of the experimental results.

This work was supported by RFBR, grant No 01-02-17398.

\newpage
\section{ Figure captions}
\begin{enumerate}
\item Energy $E$ against time-of-flight $t$ scatter plots for the right arm 
detector at $\theta_R=38^{\circ}$.
\item The missing mass $M_{dX_2}$ spectrum for
$\theta_R=38^{\circ}$.
\item The missing mass $M_{pX_1}$ spectra
obtained with CD$_2$ (the points with the statistical errors) and $^{12}$C
(the bars) targets;
(a) $\theta_R=34^{\circ}$, (b) $\theta_R=36^{\circ}$,
(c) $\theta_R=38^{\circ}$. The solid curves are normalized theoretical
predictions. The dashed curves correspond to the background interpolated by
polynomials.
\item The sum of missing mass $M_{dX_2}$ spectra 
for the reaction $pd\to pX\to pdX_2$ for the angles
$\theta_R=34^{\circ}$ and $36^{\circ}$.
\item The sum of missing mass $M_{pX_1}$ (a) and 
$M_{X_1}$ (b) spectra for the angles 
$\theta_R=34^{\circ}$ and $\theta_R=36^{\circ}$. The dashed and solid curves
are results of a fit by polynomials
for the background and Gaussians for the peaks, respectively.
\item The result of the simulation of the missing $M_{X_1}$ mass spectra;
(a) -- the proton from the decay $X\to pX_1$ is emitted in the whole allowed
angular region, (b) -- this proton is detected at 
$\theta_R=34^{\circ}$ and $\theta_R=36^{\circ}$.

\end{enumerate}

\newpage
%Fig. 1
\begin{figure}
\epsfxsize=16cm
\epsfysize=16cm
\centerline{
\epsfbox{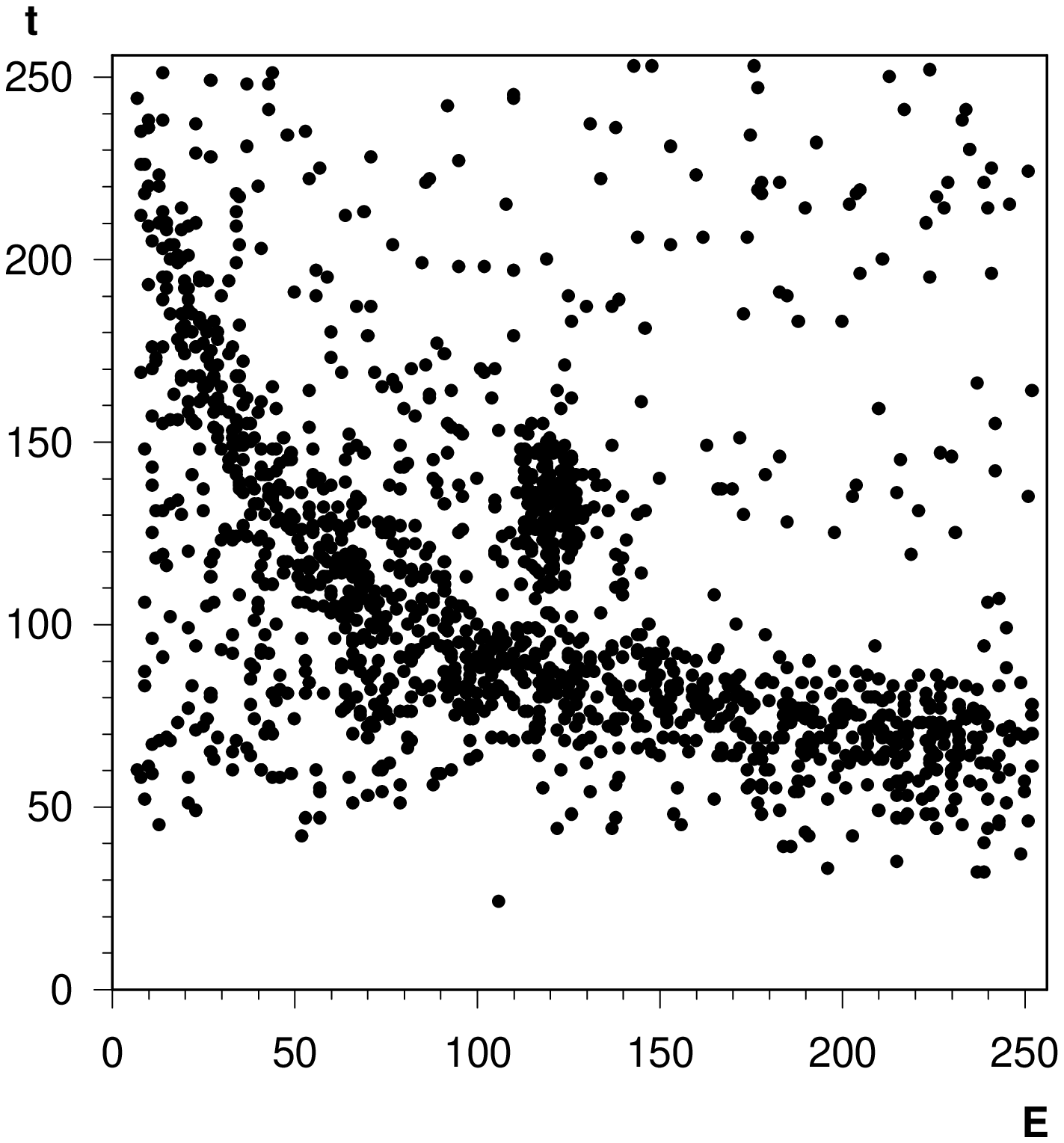}}
\caption{} 
\end{figure}

\newpage
%Fig. 2
\begin{figure}
\epsfxsize=16cm
\epsfysize=16cm
\centerline{
\epsfbox{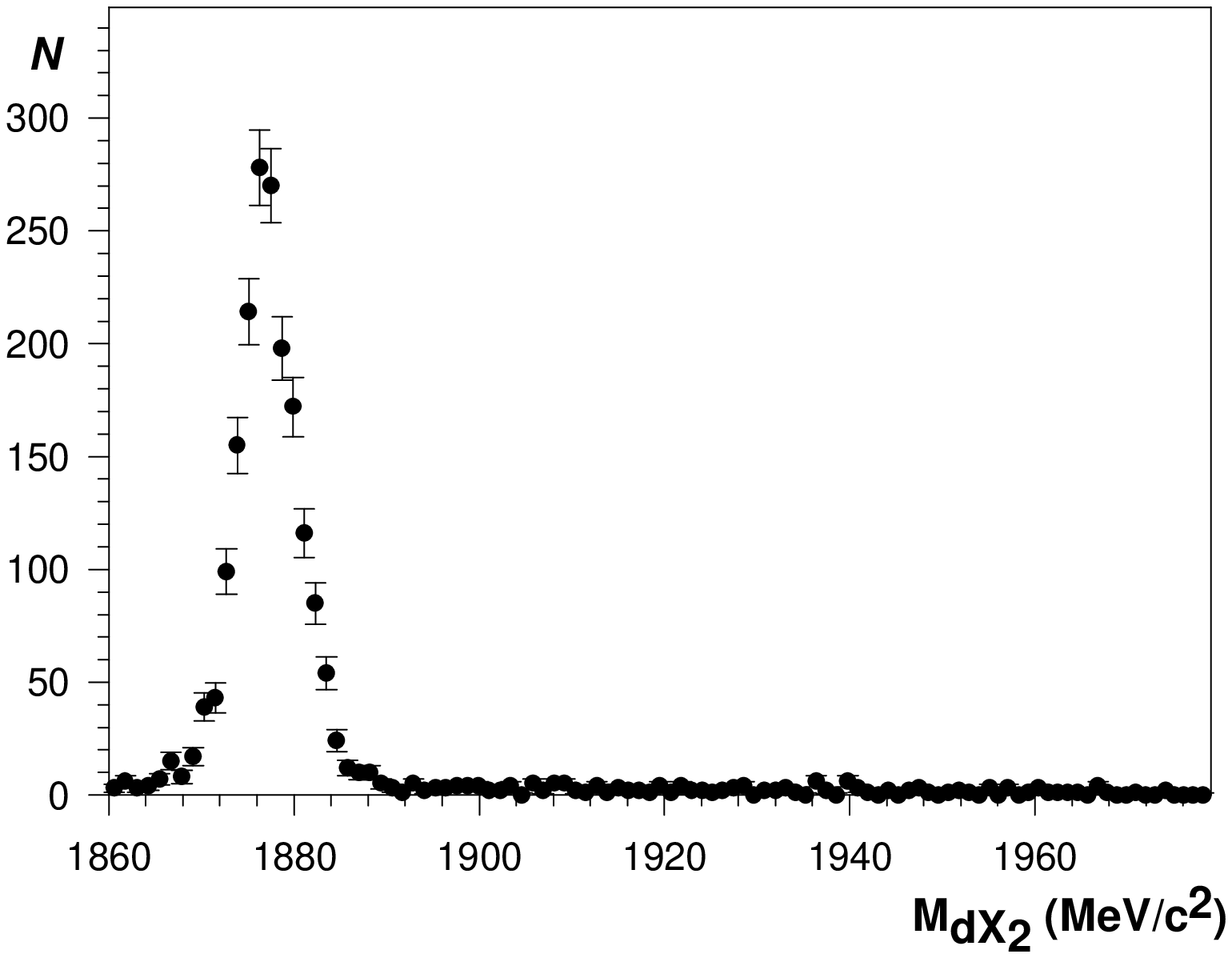}}
\caption{}
\end{figure}

\newpage
%Fig. 3
\begin{figure}
\epsfxsize=14cm
\epsfysize=18cm
\centerline{
\epsfbox{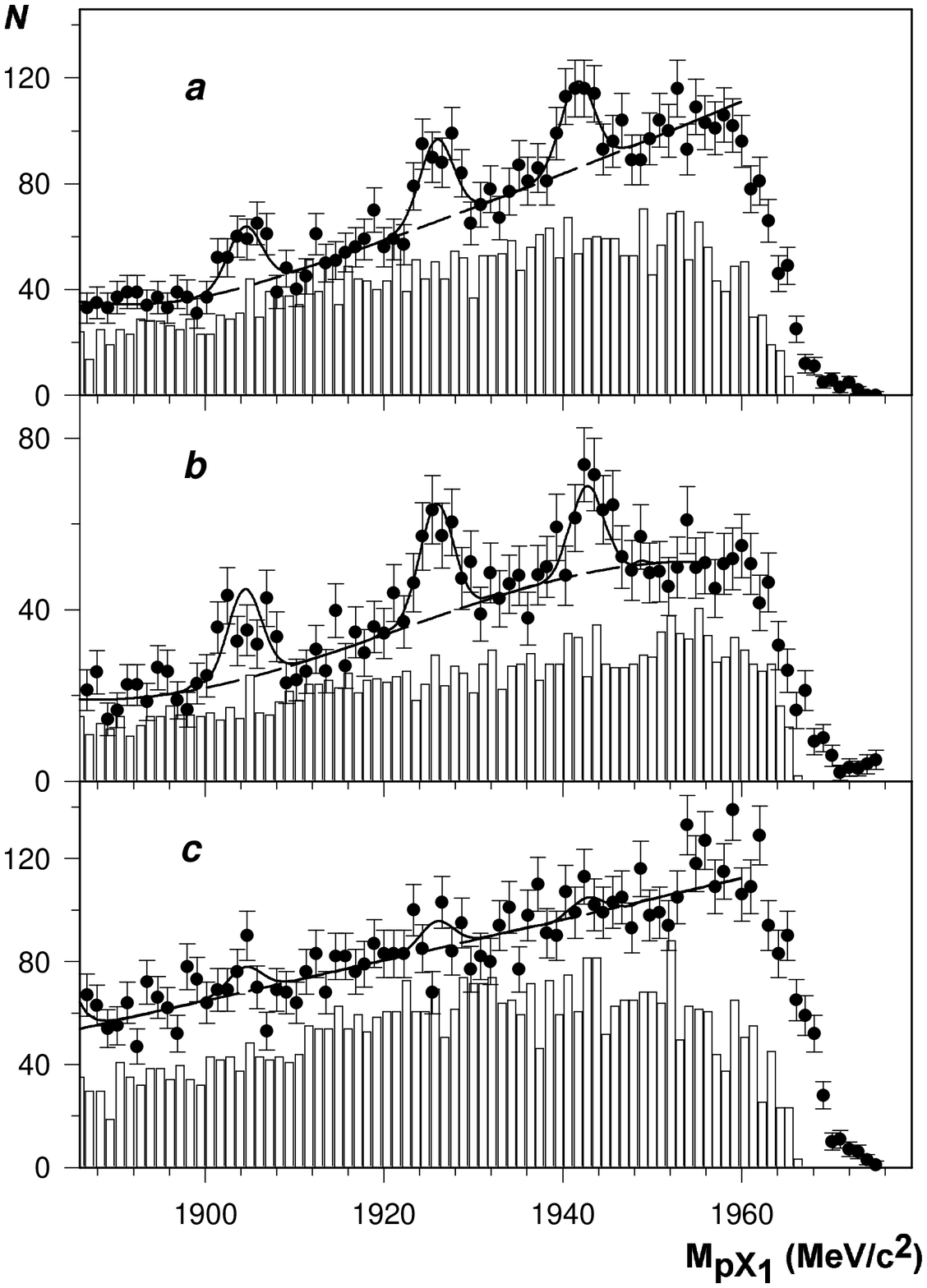}}
\caption{}
\end{figure}

\newpage
%Fig. 4
\begin{figure}
\epsfxsize=14cm
\epsfysize=16cm
\centerline{
\epsfbox{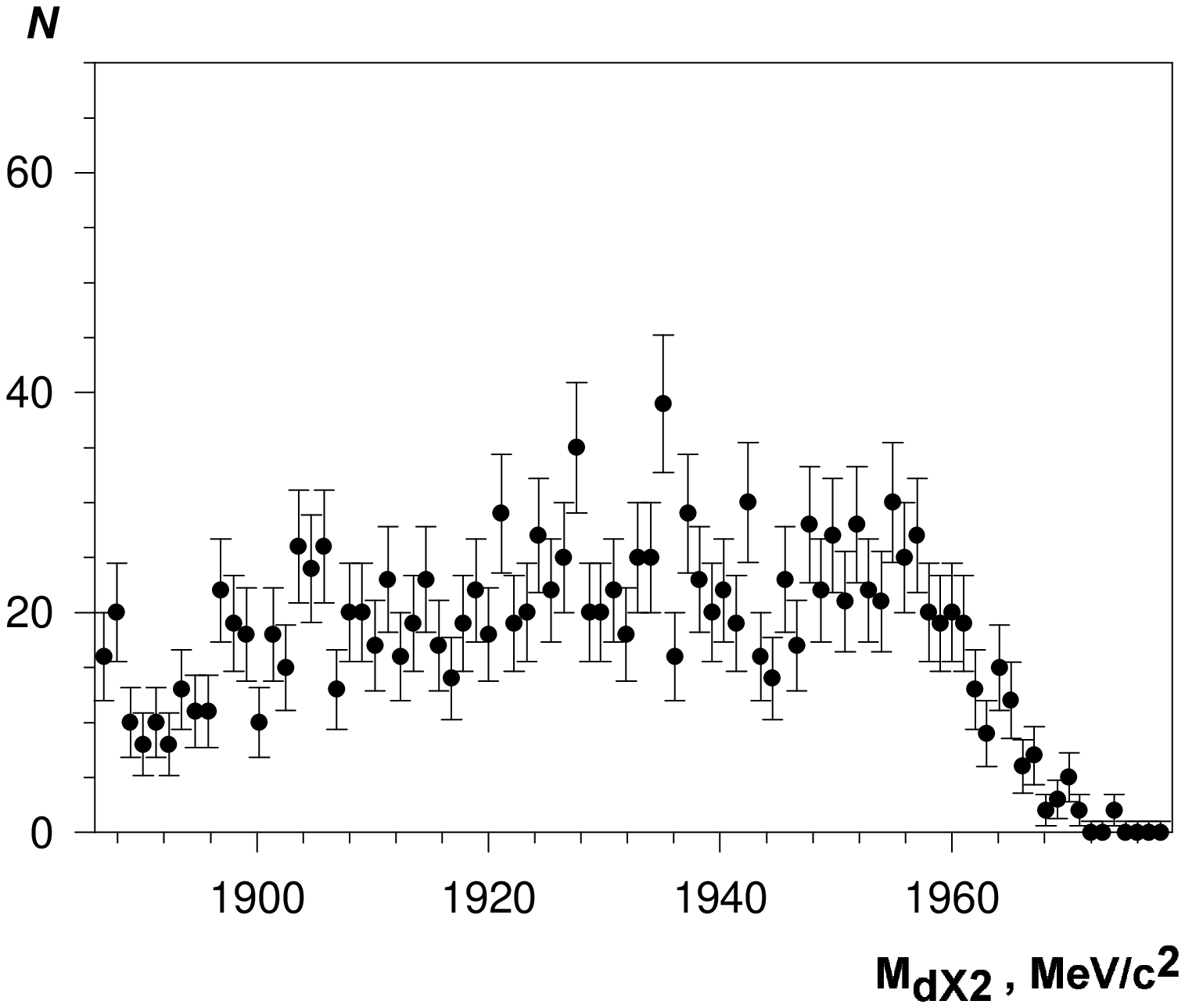}}
\caption{}
\end{figure}

\newpage
%Fig. 5
\begin{figure}
\epsfxsize=14cm
\epsfysize=16cm
\centerline{
\epsfbox{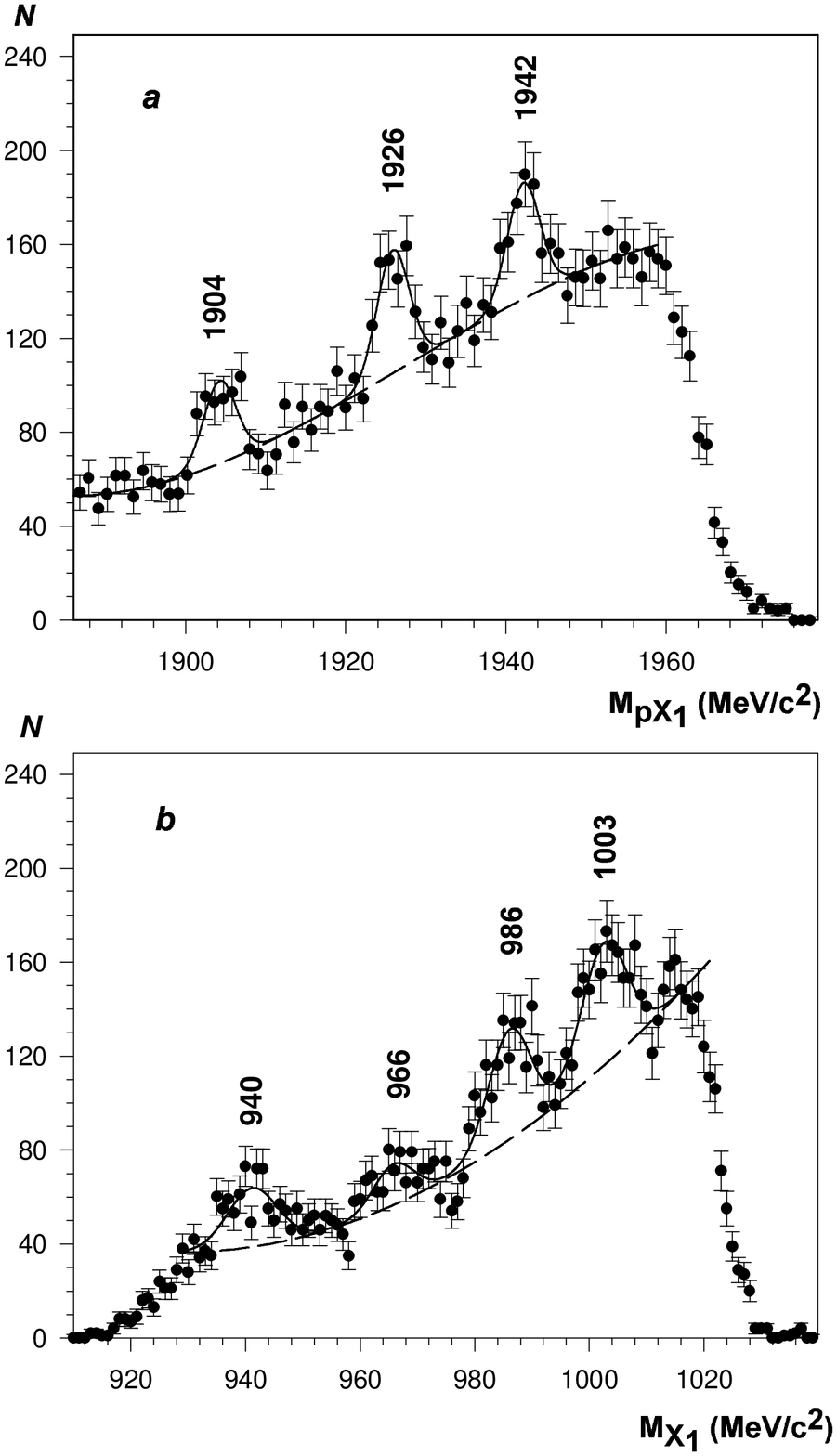}}
\caption{}
\end{figure}

\newpage
%Fig. 6
\begin{figure}
\epsfxsize=14cm
\epsfysize=16cm
\centerline{   
\epsfbox{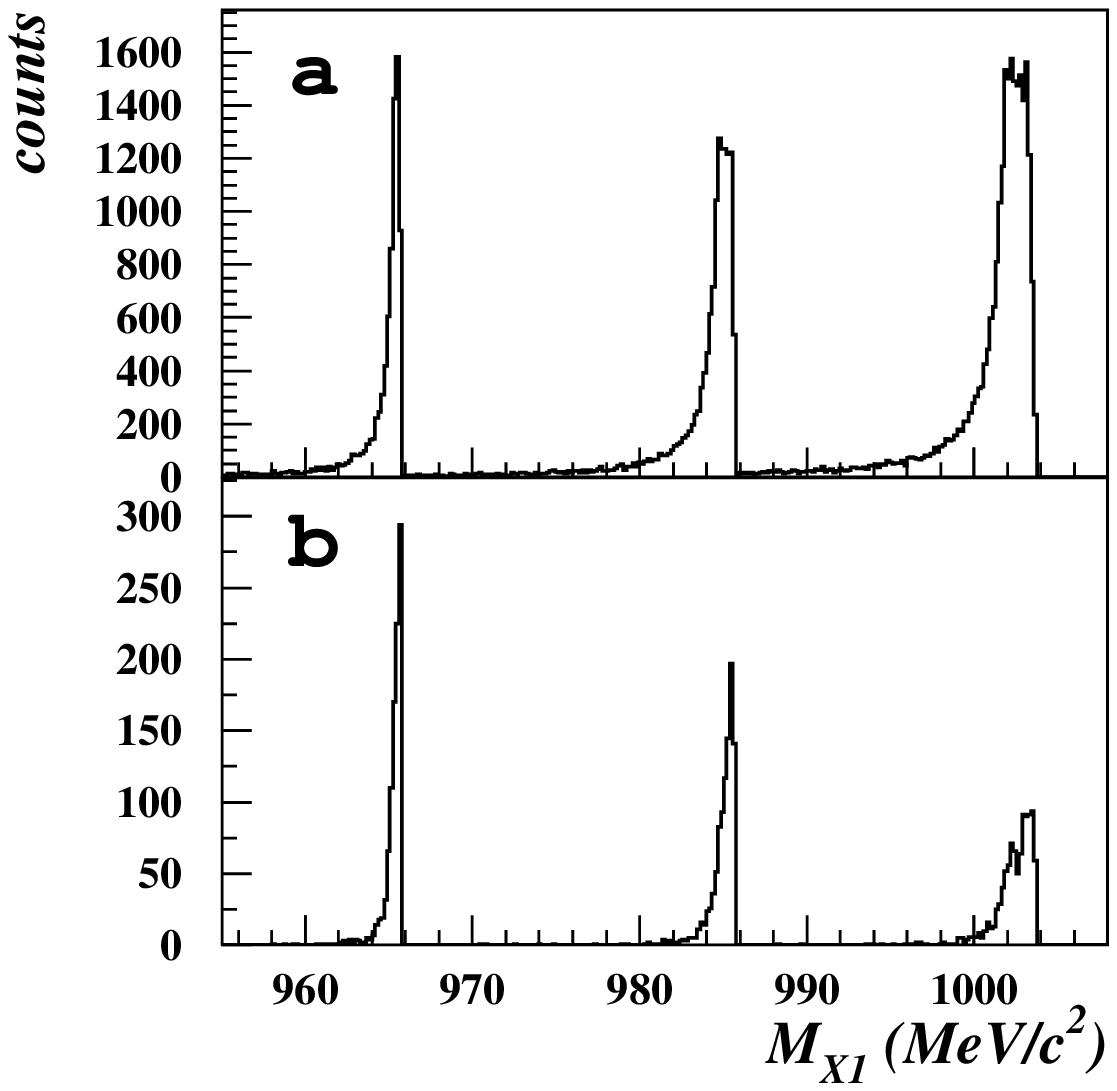}}
\caption{}
\end{figure}

\end{document}